\documentclass%
[twocolumn,showpacs,preprintnumbers,amsmath,amssymb,floatfix]{revtex4}
\usepackage{txfonts}
\usepackage{graphicx}
\usepackage{dcolumn}
\usepackage{bm}
\begin{document}
\sloppy
%
%%%% TITLE & ABSTRACT %%%%%%%%%%%%%%%%%%%%%%%%%%%%%%%%%%%%%%%%%%%%%%%%%%%%%%%%%%
%
\preprint{}

\title{\boldmath%
The $\Lambda$-$\Sigma$ coupling effect 
in the neutron-rich $\Lambda$-hypernucleus ${}_{\;\Lambda}^{10}\mathrm{Li}$
by microscopic shell model%
}
\author{A. Umeya}
\email{u-atusi@isc.osakac.ac.jp}
\author{T. Harada}
\affiliation{Research Center for Physics and Mathematics, 
Osaka Electro-Communication University, Neyagawa, Osaka, 572-8530, Japan
}%
\date{\today}
\begin{abstract}
We investigate the structure of the neutron-rich $\Lambda$-hypernucleus 
${}_{\;\Lambda}^{10}\mathrm{Li}$ 
by using microscopic shell-model calculations 
considering a $\Lambda$-$\Sigma$ coupling effect. 
The calculated $\Sigma$-mixing probability 
in the ${}_{\;\Lambda}^{10}\mathrm{Li}$ ground state 
is found to be about $0.34$ $\%$ 
which is coherently enhanced 
by the $\Lambda$-$\Sigma$ coupling configurations, 
leading to the energy shift $0.28$ $\mathrm{MeV}$ 
which is about 3 times larger than that in ${}_{\Lambda}^{7}\mathrm{Li}$. 
The importance of the $\Sigma$ configuration 
obtained by the $\Sigma N$ interaction 
and the potentiality of the neutron-rich environment 
are discussed. 
\end{abstract}
\maketitle
%
%%%% MAIN %%%%%%%%%%%%%%%%%%%%%%%%%%%%%%%%%%%%%%%%%%%%%%%%%%%%%%%%%%%%%%%%%%%%%%
%
\section{Introduction}
One of the most important subjects in strangeness nuclear physics 
is a study of neutron-rich $\Lambda$-hypernuclei~\cite{maj95}. 
It is expected that 
a $\Lambda$ hyperon 
plays a glue-like role in nuclei beyond the neutron-drip line, 
together with a strong $\Lambda N$-$\Sigma N$ coupling~\cite{swe00,aka00}, 
which might induce a $\Sigma$-mixing in nuclei. 
The knowledge of behavior of hyperons in a neutron-excess environment 
significantly affects our understanding of neutron stars, 
because it makes the Equation of State soften~\cite{bal00}. 
The purpose of our study 
is to theoretically clarify 
the structure of the neutron-rich $\Lambda$-hypernuclei 
by a nuclear shell model, 
which has successfully been applied for description of the neutron-excess 
nuclei~\cite{por08:0,ume04a,ume06a,ume08b}. 
\par
Shell-model studies for $p$-shell $\Lambda$-hypernuclei 
have been performed by several 
authors~\cite{gal71:3,dal78,mil85,aue83,ito90,mil07:2,fuj02,hal08}. 
A series of pioneering shell-model calculations 
was carried out by Gal, Soper and Dalitz \cite{gal71:3} in the 1970s, 
using $\Lambda N$ effective interactions 
parametrized from the available data. 
The $\gamma$-ray transitions for $p$-shell 
$\Lambda$-hypernuclei~\cite{dal78,mil85} 
and $\Lambda$ production cross sections 
of $(K_{}^{-}, \pi_{}^{-})$ and $(\pi_{}^{+}, K_{}^{+})$ 
reactions~\cite{aue83,ito90} 
were successfully explained by shell-model descriptions. 
Recently, 
energy spacings of $p$-shell $\Lambda$-hypernuclei 
have been studied by Millener's shell-model calculations 
including the $\Lambda$-$\Sigma$ coupling~\cite{mil07:2}, 
so as to interpret the precise data 
of $p$-shell \mbox{$N \approx Z$} $\Lambda$-hypernuclei 
from $\gamma$-ray measurements by $\mathrm{Ge}$ detector~\cite{tam08:0}. 
Moreover, 
properties of the $\Lambda N$ effective interactions 
derived form the recent Nijmegen potentials~\cite{mae89,rij99,rij06} 
in shell-model calculations have been discussed~\cite{fuj02,hal08}. 
\par
Recently, 
Saha \textit{et~al.}\ 
have performed the first successful measurement 
of a neutron-rich $\Lambda$-hypernucleus ${}_{\;\Lambda}^{10}\mathrm{Li}$ 
by the double-charge exchange reaction $(\pi_{}^{-}, K_{}^{+})$ 
on a ${}_{}^{10}\mathrm{B}$ target~\cite{sah05}. 
However, 
the magnitude and incident-momentum dependence 
of the experimental production cross sections 
cannot be reproduced in a theoretical calculation 
by Tretyakova and Lanskoy~\cite{tre03}. 
They predicted that the cross section for ${}_{\;\Lambda}^{10}\mathrm{Li}$ 
is mainly explained by a two-step process, 
$\pi_{}^{-} p$ $\to$ $K_{}^{0} \Lambda$ 
followed by 
$K_{}^{0} p$ $\to$ $K_{}^{+} n$, 
or 
$\pi_{}^{-} p$ $\to$ $\pi_{}^{0} n$ 
followed by 
$\pi_{}^{0} p$ $\to$ $K_{}^{+} \Lambda$ 
with the distorted-wave impulse approximation, 
rather than by a one-step process, 
$\pi_{}^{-} p$ $\to$ $K_{}^{+} \Sigma_{}^{-}$ 
via $\Sigma_{}^{-} p$ doorways due to the
$\Sigma_{}^{-} p$ $\leftrightarrow$ $\Lambda n$ 
coupling. 
This problem might suggest the importance of a $\Sigma$-mixing 
in the $\Lambda$-hypernucleus. 
The analysis of the ${}_{}^{10}\mathrm{B}$ $(\pi_{}^{-}, K_{}^{+})$ reaction 
provides a reason to carefully examine wave functions 
involving $\Sigma$ admixtures in ${}_{\;\Lambda}^{10}\mathrm{Li}$, 
as well as a mechanism of this reaction~\cite{har09}. 
\par
In this paper, 
we investigate the structure 
of the neutron-rich hypernucleus ${}_{\;\Lambda}^{10}\mathrm{Li}$, 
in microscopic shell-model calculations 
considering the $\Lambda$-$\Sigma$ coupling effect. 
We focus on the $\Sigma$-mixing probabilities and the energy shifts 
of the neutron-rich hypernucleus. 
Also, 
we investigate the effect of the $\Sigma N$ interaction 
to the core-nuclear state by the perturbation method. 
We discuss that 
the coupling strengths 
are enhanced in the neutron-rich $\Lambda$-hypernucleus 
and 
are related to the $\beta$-transition properties of the nuclear core state. 
%
%%%%%%%%%%%%%%%%%%%%%%%%%%%%%%%%%%%%%%%%%%%%%%%%%%%%%%%%%%%%%%%%%%%%%%%%%%%%%%%%
\section{Formalism}
\label{form}
%%%%%%%%%%%%%%%%%%%%%%%%%%%%%%%%%%%%%%%%%%%%%%%%%%%%%%%%%%%%%%%%%%%%%%%%%%%%%%%%
%
\subsection{\boldmath Multi-configuration shell model for $\Lambda$-hypernuclei}
\label{1-form}
We consider a $\Lambda$-nuclear state 
involving a $\Sigma$-mixing in a $\Lambda$-hypernucleus ${}_{\Lambda}^{A}Z$ 
with the mass number $A$ and the atomic number $Z$ 
in a multi-configuration nuclear shell model. 
The state of the $\Lambda$-hypernucleus 
is represented by $| ({}_{\Lambda}^{A} Z) \nu T T_{z}^{} J M \rangle$, 
where $T$ and $T_{z}^{}$ are the isospin and its $z$-component, 
respectively, 
and $J$ and $M$ for the angular momentum. 
The index $\nu$ is introduced to distinguish states with the same $T$ and $J$. 
\par
In the configuration space for the $\Lambda$-hypernucleus 
involving a $\Lambda$-$\Sigma$ coupling, 
the Hamiltonian is given as 
\begin{gather}
\begin{aligned}[b]
  H
  &=
  H_{\Lambda}^{}
  +
  H_{\Sigma}^{}
  +
  V_{\Lambda\Sigma}^{}
  +
  V_{\Sigma\Lambda}^{}
  ,
\end{aligned}
\label{e01-form}
\end{gather}
where 
$H_{\Lambda}^{}$ is the Hamiltonian in the $\Lambda$ configuration space 
and 
$H_{\Sigma}^{}$ is that in the $\Sigma$ configuration space. 
$V_{\Lambda\Sigma}^{}$ and its Hermitian conjugate $V_{\Sigma\Lambda}^{}$ 
denote the two-body $\Lambda$-$\Sigma$ coupling interaction, 
$\Lambda N$ $\leftrightarrow$ $\Sigma N$. 
Then, 
we can write the $\Lambda$-nuclear state with $T$, $J$ as 
\begin{gather}
\begin{aligned}[b]
  | ({}_{\Lambda}^{A} Z) \nu T J \rangle
  &=
  \sum_{\mu}^{} 
  C_{\nu,\mu}^{}
  | \psi_{\mu}^{\Lambda}; T J \rangle
  +
  \sum_{\mu_{}^{\prime}}^{} 
  D_{\nu,\mu_{}^{\prime}}^{}
  | \psi_{\mu_{}^{\prime}}^{\Sigma}; T J \rangle
  ,
\end{aligned}
\label{e02-form}
\end{gather}
where 
$| \psi_{\mu}^{\Lambda}; T J \rangle$ and 
$| \psi_{\mu_{}^{\prime}}^{\Sigma}; T J \rangle$ 
are eigenstates 
for the $\Lambda$ and $\Sigma$ configurations, 
respectively, 
which are given by 
\begin{gather}
\begin{aligned}[b]
  H_{\Lambda}^{}
  | \psi_{\mu}^{\Lambda}; T J \rangle
  &=
  E_{\mu}^{\Lambda}
  | \psi_{\mu}^{\Lambda}; T J \rangle
  ,
\end{aligned}
\label{e03-form}
\\
\begin{aligned}[b]
  H_{\Sigma}^{}
  | \psi_{\mu_{}^{\prime}}^{\Sigma}; T J \rangle
  &=
  E_{\mu_{}^{\prime}}^{\Sigma}
  | \psi_{\mu_{}^{\prime}}^{\Sigma}; T J \rangle
  .
\end{aligned}
\label{e04-form}
\end{gather}
Although the coefficients $C_{\nu,\mu}^{}$ and $D_{\nu,\mu_{}^{\prime}}^{}$ 
are determined by diagonalization of the full Hamiltonian $H$, 
we treat $V_{\Lambda\Sigma}^{}$ and $V_{\Sigma\Lambda}^{}$ as perturbation 
because a $\Sigma$ hyperon has a larger mass 
than a $\Lambda$ hyperon by about $80$ $\mathrm{MeV}$\@. 
When taking into account up to the first-order terms, 
the coefficients can be written as 
\begin{gather}
\begin{aligned}[b]
  C_{\nu,\mu}^{}
  &=
  \delta_{\nu \mu}^{}
  ,
\end{aligned}
\label{e05-form}
\\
\begin{aligned}[b]
  D_{\nu,\mu_{}^{\prime}}^{}
  &=
  -
  \frac{ \langle \psi_{\nu}^{\Lambda}; T J |
         V_{\Lambda\Sigma}^{}
         | \psi_{\mu_{}^{\prime}}^{\Sigma}; T J \rangle }
       { E_{\mu_{}^{\prime}}^{\Sigma} - E_{\nu}^{\Lambda} }
  .
\end{aligned}
\label{e06-form}
\end{gather}
Since a $\Lambda$-$\Sigma$ coupling strength 
for each $\Sigma$ eigenstate 
$| \psi_{\mu_{}^{\prime}}^{\Sigma}; T J \rangle$ 
is obtained as $| D_{\nu,\mu_{}^{\prime}}^{} |_{}^{2}$, 
the $\Sigma$-mixing probability 
in the $\Lambda$-nuclear state $| ({}_{\Lambda}^{A} Z) \nu T J \rangle$ 
is given as 
\begin{gather}
\begin{aligned}[b]
  P_{\Sigma}^{}
  &=
  \sum_{\mu_{}^{\prime}}^{}
  | D_{\nu,\mu_{}^{\prime}}^{} |_{}^{2}
\end{aligned}
\label{e07-form}
\end{gather}
or 
\begin{gather}
\begin{aligned}[b]
  P_{\Sigma}^{(R)}
  &=
  \mathinner{\biggl(
    \sum_{\mu_{}^{\prime}}^{}
    | D_{\nu,\mu_{}^{\prime}}^{} |_{}^{2}
  \biggr)}
  \biggm/
  \mathinner{\biggl(
    1
    +
    \sum_{\mu_{}^{\prime}}^{}
    | D_{\nu,\mu_{}^{\prime}}^{} |_{}^{2}
  \biggr)}
  ,
\end{aligned}
\label{e08-form}
\end{gather}
where the latter probability is a renormalized value. 
The binding energy $E_{\nu}^{\Lambda} - \Delta E$ 
is calculated by the energy shift 
which is given as 
\begin{gather}
\begin{aligned}[b]
  \Delta E
  &=
  \sum_{\mu_{}^{\prime}}^{}
  \left( E_{\mu_{}^{\prime}}^{\Sigma} - E_{\nu}^{\Lambda} \right)
  | D_{\nu,\mu_{}^{\prime}}^{} |_{}^{2}
\end{aligned}
\label{e09-form}
\end{gather}
in the perturbation theory. 
\par
It is well known that 
a $\Lambda$ hyperon in a $\Lambda$-hypernucleus 
is described by the single-particle picture very well 
because the $\Lambda N$ interaction is weak. 
On the other hand, 
in terms of a $\Sigma$ hyperon, 
the nuclear configuration 
would change due to the strong spin-isospin dependence 
in the $\Sigma N$ interaction~\cite{dov89}. 
In order to evaluate the single-particle picture for a hyperon, 
we consider a spectroscopic factor for a hyperon-pickup from 
$| \psi_{\mu}^{Y}; T J \rangle$, 
\begin{gather}
\begin{aligned}[b]
  S_{\mu}^{}( \nu_{N}^{} T_{N}^{} J_{N}^{}, j_{Y}^{} )
  &=
  \frac{ \left|
           \langle \psi_{\mu}^{Y}; T J \|
           \bm{a}_{ j_{Y}^{} }^{\dagger}
           \| ({}_{}^{A-1} Z) \nu_{N}^{} T_{N}^{} J_{N}^{} \rangle
         \right|_{}^{2} }
       { \mathinner{(2T+1)} \mathinner{(2J+1)} }
  ,
\end{aligned}
\label{e10-form}
\end{gather}
where 
$| ({}_{}^{A-1} Z) \nu_{N}^{} T_{N}^{} J_{N}^{} \rangle$ 
is an eigenstate of the core nucleus, 
which is obtained by diagonalizing in the nucleon configurations, 
and 
$\bm{a}_{ j_{Y}^{} }^{\dagger}$ 
is a creation operator of a single-particle state of the hyperon 
in an orbit $j_{Y}^{}$. 
The matrix element 
$\langle \cdot \| \cdot \| \cdot \rangle$ 
with the Edmonds' convention~\cite{edm60} 
is reduced with respect to both the isospin and the angular momentum. 
The spectroscopic factor satisfies the sum rule 
\begin{gather}
\begin{aligned}[b]
  \sum_{ \nu_{N}^{} T_{N}^{} J_{N}^{} }^{}
  S_{\mu}^{}( \nu_{N}^{} T_{N}^{} J_{N}^{}, j_{Y}^{} )
  &=
  n_{ j_{Y}^{} }^{}
  ,
\end{aligned}
\label{e11-form}
\end{gather}
where $n_{ j_{Y}^{} }^{}$ is the number of hyperons in the orbit $j_{Y}^{}$. 
If a hyperon in the hypernucleus provides the single-particle nature, 
the state $| \psi_{\mu}^{Y}; T J \rangle$ 
is represented as a tensor product of 
a core-nuclear state 
$| ({}_{}^{A-1}Z) \nu_{\mathrm{core}}^{} T_{N}^{} J_{N}^{} \rangle$ 
and a hyperon state $| j_{Y}^{} \rangle$; 
we obtain 
\mbox{$S_{\mu}^{}( \nu_{N}^{} T_{N}^{} J_{N}^{}, j_{Y}^{} )
       = \delta_{\nu_{N}^{} \nu_{\mathrm{core}}^{}}^{}$}, 
where \mbox{$\nu_{N}^{} = \nu_{\mathrm{core}}^{}$} means 
the core state 
is equivalent to the ${}_{}^{A-1}Z$ state in the weak-coupling limit. 
%
%%%%%%%%%%%%%%%%%%%%%%%%%%%%%%%%%%%%%%%%%%%%%%%%%%%%%%%%%%%%%%%%%%%%%%%%%%%%%%%%
%
\subsection{Shell-model setup and effective interactions}
\label{2-form}
In the present shell-model calculations, 
we construct wave functions of ${}_{\Lambda}^{A}Z$ as follows: 
Four nucleons are inert in the ${}_{}^{4}\mathrm{He}$ core 
and 
\mbox{$(A-5)$} valence nucleons move in the $p$-shell orbits. 
The $\Lambda$ or $\Sigma$ hyperon 
is assumed to be in the lowest $0s_{1/2}^{}$ orbit. 
For the $NN$ effective interaction, 
we adopt the Cohen-Kurath (8--16) 2BME~\cite{coh65}, 
which is a traditional and empirical interaction for ordinary $p$-shell nuclei, 
and 
is one of the reliable effective interactions 
for stable and semi-stable nuclei. 
The $YN$ effective interaction 
is written~\cite{gal71:3,dal78,mil85} as 
\begin{gather}
\begin{aligned}[b]
  V_{Y}^{}
  &=
  V_{0}^{}(r)
  +
  V_{\sigma}^{}(r)
  \bm{s}_{N}^{} \cdot \bm{s}_{Y}^{}
  +
  V_{\mathrm{LS}}^{}(r)
  \bm{\ell} \cdot \mathinner{( \bm{s}_{N}^{} + \bm{s}_{Y}^{} )}
\\
  & \hspace{1em}
  {}
  +
  V_{\mathrm{ALS}}^{}(r)
  \bm{\ell} \cdot \mathinner{( - \bm{s}_{N}^{} + \bm{s}_{Y}^{} )}
  +
  V_{T}^{}(r)
  S_{12}^{}
  ,
\end{aligned}
\label{e12-form}
\end{gather}
where 
$V(r)$'s are radial functions of the relative coordinate 
\mbox{$r = | \bm{r}_{N}^{} - \bm{r}_{Y}^{} |$} 
between the nucleon and the hyperon. 
$\bm{s}_{N}^{}$ and $\bm{s}_{Y}^{}$ 
are spin operators for the nucleon and the hyperon, 
respectively, 
and 
$\bm{\ell}$ is the angular momentum operator of the relative motion. 
The tensor operator $S_{12}^{}$ is defined by 
\begin{gather}
\begin{aligned}[b]
  S_{12}^{}
  &=
  3
  \mathinner{( \hat{\bm{r}} \cdot \bm{\sigma}_{N}^{} )}
  \mathinner{( \hat{\bm{r}} \cdot \bm{\sigma}_{Y}^{} )}
  -
  \mathinner{( \bm{\sigma}_{N}^{} \cdot \bm{\sigma}_{Y}^{} )}
\end{aligned}
\label{e13-form}
\end{gather}
with \mbox{$\bm{\sigma} = 2 \bm{s}$} 
and \mbox{$\hat{\bm{r}} = (\bm{r}_{N}^{} - \bm{r}_{Y}^{}) / r$}. 
In Table~\ref{t01}, 
we list the parameters of radial integrals 
$\bar{V}$, $\Delta$, $S_{+}^{}$, $S_{-}^{}$ and $T$, 
which correspond to 
$V_{0}^{}$, $V_{\sigma}^{}$, $V_{\mathrm{LS}}^{}$, 
$V_{\mathrm{ALS}}^{}$ and $V_{T}^{}$, 
respectively. 
The parameters $S_{+}^{}$ and $S_{-}^{}$ 
are defined as the coefficients of 
$\bm{\ell}_{N}^{} \cdot \mathinner{( \bm{s}_{N}^{} + \bm{s}_{Y}^{} )}$ 
and 
$\bm{\ell}_{N}^{} \cdot \mathinner{( - \bm{s}_{N}^{} + \bm{s}_{Y}^{} )}$, 
respectively, 
where 
$\bm{\ell}_{N}^{}$ is the angular momentum operator of the nucleon 
and is proportional to the relative $\bm{\ell}$ 
for the hyperon in the $0s_{1/2}^{}$ orbit. 
We adopt the values of $V_{\Lambda}^{}$ 
which is given in Ref.~\cite{mil07:2}. 
For $V_{\Lambda\Sigma}^{}$ and $V_{\Sigma}^{}$, 
we used the $\Lambda N$-$\Sigma N$ and $\Sigma N$ effective 
interactions~\cite{mil07:2,mil:p} 
based on the NSC97e,f potentials~\cite{rij99}. 
\begin{table}[t]
\caption[$YN$ interactions]{%
Radial integrals for $YN$ effective interactions 
in unit of $\mathrm{MeV}$\@. 
The values are listed in Ref.~\cite{mil07:2} 
for the $\Lambda N$ interaction $V_{\Lambda}^{}$
and the $\Lambda$-$\Sigma$ coupling interaction $V_{\Lambda\Sigma}^{}$, 
and Ref.~\cite{mil:p} for the $\Sigma N$ interaction $V_{\Sigma}^{}$.%
}%
\label{t01}
\begin{ruledtabular}
\begin{tabular}{lcrrrrr}
  & Isospin
  & \multicolumn{1}{c}{$\bar{V}$}
  & \multicolumn{1}{c}{$\Delta$}
  & \multicolumn{1}{c}{$S_{+}^{}$}
  & \multicolumn{1}{c}{$S_{-}^{}$}
  & \multicolumn{1}{c}{$T$}
\\ \hline 
  $V_{\Lambda}^{}$ 
  & $T=\frac{1}{2}$ 
  & $-1.2200$ & $ 0.4300$ & $-0.2025$ & $ 0.1875$ & $ 0.0300$ \\[0.5ex]
  $V_{\Sigma}^{}$ 
  & $T=\frac{1}{2}$ 
  & $ 1.0100$ & $-7.2150$ & $-0.0010$ & $ 0.0000$ & $-0.3640$ \\[0.5ex]
  $V_{\Sigma}^{}$ 
  & $T=\frac{3}{2}$ 
  & $-1.1070$ & $ 2.2750$ & $-0.2680$ & $ 0.0000$ & $ 0.1870$ \\[0.5ex]
  $V_{\Lambda\Sigma}^{}, V_{\Sigma\Lambda}^{}$
  & $T=\frac{1}{2}$ 
  & $ 1.4500$ & $ 3.0400$ & $-0.0850$ & $ 0.0000$ & $ 0.1570$
\end{tabular}
\end{ruledtabular}
\end{table}
%
%%%%%%%%%%%%%%%%%%%%%%%%%%%%%%%%%%%%%%%%%%%%%%%%%%%%%%%%%%%%%%%%%%%%%%%%%%%%%%%%
\section{Numerical results}
\label{rd}
%%%%%%%%%%%%%%%%%%%%%%%%%%%%%%%%%%%%%%%%%%%%%%%%%%%%%%%%%%%%%%%%%%%%%%%%%%%%%%%%
%
\subsection{\boldmath $\Sigma$-mixing probabilities and energy shifts}
\label{1-rd}
We perform numerical calculations 
of the neutron-rich $\Lambda$-hypernucleus ${}_{\;\Lambda}^{10}\mathrm{Li}$ 
in order to evaluate the $\Sigma$-mixing probabilities and the energy shifts 
which are obtained by Eqs.~(\ref{e07-form})--(\ref{e09-form}). 
In order to check our shell-model calculation, 
we compare our numerical results 
for the \mbox{$Z \approx N$} $\Lambda$-hypernuclei 
to other work~\cite{mil07:2}. 
We obtain the energy shifts, 
e.g., 
\mbox{$\Delta E = 0.085$} and $0.073$ $\mathrm{MeV}$ 
for the ground states of 
${}_{\Lambda}^{7}\mathrm{Li}$ and ${}_{\;\Lambda}^{11}\mathrm{B}$, 
which are comparable to the Millener's results 
of $0.078$ and $0.066$ $\mathrm{MeV}$, 
respectively. 
Therefore, 
we confirm that our calculations fully reproduce the Millener's results. 
\par
\begin{figure}[t]
  \includegraphics*[width=0.9\linewidth]{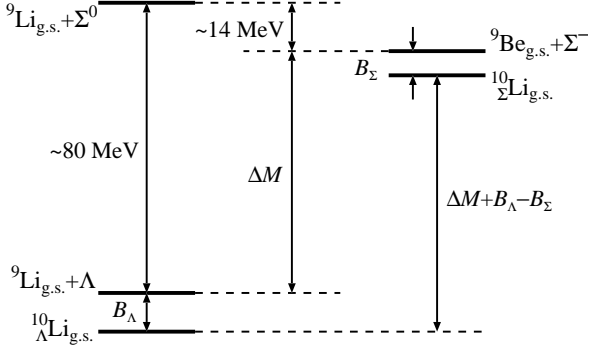}
  \caption%
  [Schematic energy levels]%
  {Schematic energy levels 
   for $\Lambda$ and $\Sigma$ ground states of ${}_{\;Y}^{10}\mathrm{Li}$.}
\label{f01}
\end{figure}

In Fig.~\ref{f01}, 
we show the schematic energy levels 
for the $\Lambda$ and $\Sigma$ ground states of ${}_{\;Y}^{10}\mathrm{Li}$. 
Here, 
we assume that 
the difference between $\Lambda$ and $\Sigma$ threshold energies 
is 
$E( {}_{}^{9}\mathrm{Li}_{\mathrm{g.s.}}^{} {+} \Sigma ) 
 - 
 E( {}_{}^{9}\mathrm{Li}_{\mathrm{g.s.}}^{} {+} \Lambda ) = 80$ 
$\mathrm{MeV}$. 
Thus the energy of the $\Sigma$ ground state 
$| \psi_{\mathrm{g.s.}}^{\Sigma} \rangle$ 
is calculated to be 
$E_{\mathrm{g.s.}}^{\Sigma} - E_{\mathrm{g.s.}}^{\Lambda}
 =
 \Delta M + B_{\Lambda}^{} - B_{\Sigma}^{}
 =
 69.3$ 
$\mathrm{MeV}$, 
measured from that of the $\Lambda$ ground state 
$| \psi_{\mathrm{g.s.}}^{\Lambda} \rangle$. 
\par
\begin{figure}[t]
  \includegraphics*[width=0.9\linewidth]{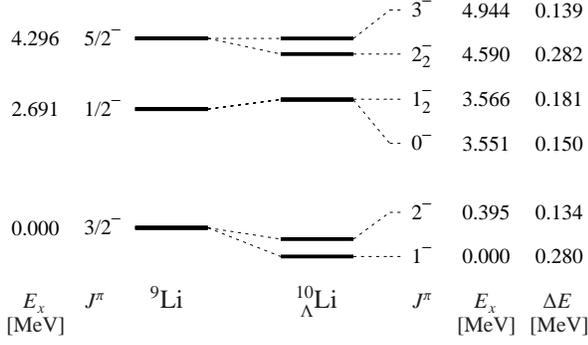}
  \caption%
  [The energy spectra of ${}_{\;\Lambda}^{10}\mathrm{Li}$]%
  {The energy spectra of ${}_{\;\Lambda}^{10}\mathrm{Li}$ 
   by the shell-model calculation with the $\Lambda$-$\Sigma$ coupling. 
   The experimental ${}_{}^{9}\mathrm{Li}$ binding energies 
   are taken from Ref.~\cite{til04}. 
   All states in the figure have the isospin \mbox{$T=\frac{3}{2}$}.}
\label{f02}
\end{figure}
The calculated energy levels for the ground and low-lying excited states 
in ${}_{\;\Lambda}^{10}\mathrm{Li}$ 
are shown in Fig.~\ref{f02}. 
Here, 
the experimental ${}_{}^{9}\mathrm{Li}$ binding energies~\cite{til04}
are used for the core-nucleus energies, 
instead of the calculated ones. 
The energy shift $\Delta E$ for ${}_{\;\Lambda}^{10}\mathrm{Li}$ 
is obtained to be $0.280$ $\mathrm{MeV}$ 
which is about 3 times larger than 
that for ${}_{\Lambda}^{7}\mathrm{Li}$ or ${}_{\;\Lambda}^{11}\mathrm{B}$. 
As shown in Fig.~\ref{f02}, 
the values of energy shifts for the low-lying excited states 
also account for about a few hundreds of $\mathrm{keV}$. 
In particular, 
the second $2_{}^{-}$ state with the $4.604$ $\mathrm{MeV}$ excitation energy 
has the large energy shift of $0.282$ $\mathrm{MeV}$. 
The calculated $\Sigma$-mixing probabilities 
are shown in Table~\ref{t02}. 
It is found that 
the $\Sigma$-mixing probability is 
\mbox{$P_{\Sigma}^{} = 0.345$} $\%$ 
in the $1_{\mathrm{g.s.}}^{-}$ ground state 
of ${}_{\;\Lambda}^{10}\mathrm{Li}$, 
where the effect of renormalization is very small, 
\mbox{$P_{\Sigma}^{(R)} = 0.344$} $\%$. 
%%%%%%%%%%%%%%%%%%%%%%%%%%%%%%
The $\Sigma_{}^{-}$ and $\Sigma_{}^{0}$ admixtures 
for the ${}_{\;\Lambda}^{10}\mathrm{Li}$ ground state 
are 
$P_{\Sigma_{}^{-}}^{} = 0.183$ $\%$ and $P_{\Sigma_{}^{0}}^{} = 0.159$ $\%$, 
respectively. 
The $\Sigma_{}^{+}$ admixture is negligible. 
%%%%%%%%%%%%%%%%%%%%%%%%%%%%%%
This is the first result that 
the $\Sigma$-mixing probabilities and the energy shifts 
in the neutron-rich hypernucleus 
are coherently enhanced by the configuration mixing 
in the nuclear shell model. 
\begin{table}[t]
\caption[$\Sigma$-mixing probabilities]{%
The calculated $\Sigma$-mixing probabilities of 
${}_{\;\Lambda}^{10}\mathrm{Li}$. 
All states in the table have the isospin $T=3/2$. 
}%
\label{t02}
\begin{ruledtabular}
\begin{tabular}{lcccc}
    $J_{}^{\pi}$
  & $P_{\Sigma}^{}$ $[\%]$
  & $P_{\Sigma_{}^{-}}^{}$ $[\%]$
  & $P_{\Sigma_{}^{0}}^{}$ $[\%]$
  & $P_{\Sigma_{}^{+}}^{}$ $[\%]$
\\ \hline
  $1_{\mathrm{g.s.}}^{-}$ &  $0.345$  &  $0.183$  &  $0.159$  &  $0.002$  \\
  $2_{}^{-}$              &  $0.166$  &  $0.096$  &  $0.070$  &  $0.000$  \\
  $0_{}^{-}$              &  $0.185$  &  $0.098$  &  $0.086$  &  $0.001$  \\
  $1_{2}^{-}$             &  $0.227$  &  $0.128$  &  $0.098$  &  $0.001$  \\
  $2_{2}^{-}$             &  $0.350$  &  $0.187$  &  $0.162$  &  $0.001$  \\
  $3_{}^{-}$              &  $0.175$  &  $0.104$  &  $0.071$  &  ---
\end{tabular}
\end{ruledtabular}
\end{table}
%
%%%%%%%%%%%%%%%%%%%%%%%%%%%%%%%%%%%%%%%%%%%%%%%%%%%%%%%%%%%%%%%%%%%%%%%%%%%%%%%%
%
\subsection{\boldmath$\Lambda$ and $\Sigma$ eigenstates}
\label{2-rd}
We examine an enhancement of $\Sigma$-mixing probabilities and energy shifts 
of ${}_{\;\Lambda}^{10}\mathrm{Li}$ eigenstates 
by the configuration mixing in the $\Sigma$ states, 
which couple to the $\Lambda$ states 
by the $\Lambda$-$\Sigma$ coupling. 
In order to investigate a change of the core-nuclear configuration 
due to the $YN$ interaction, 
we estimate the energy spectra and the hyperon-pickup spectroscopic factors 
in the $\Lambda$ and $\Sigma$ eigenstates 
with \mbox{$T=\frac{3}{2}$}, \mbox{$J_{}^{\pi}=1_{}^{-}$} 
including the ground and excited states. 
\par
\begin{figure}[t]
  \includegraphics*[width=\linewidth]{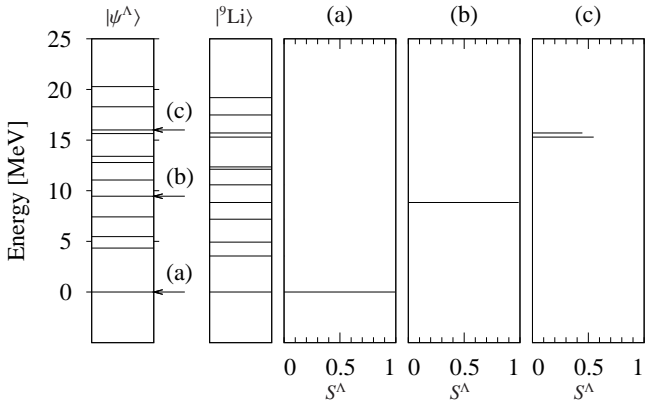}
  \caption%
  [Calculated energy spectra and spectroscopic factors]%
  {Calculated energy spectra 
   for $\Lambda$ eigenstates of ${}_{\;\Lambda}^{10}\mathrm{Li}$ 
   and eigenstates of ${}_{}^{9}\mathrm{Li}$. 
   $\Lambda$-pickup spectroscopic factors for three eigenstates, 
   labeled by (a), (b), and (c), 
   are shown in each panel.} 
\label{f03}
\end{figure}

The calculated energy spectra of $\Lambda$ eigenstates 
$| \psi_{\mu}^{\Lambda}; T J \rangle$ 
are shown in the left panel of Fig.~\ref{f03}, 
measured from the ground state. 
We find that 
the energy spacings between the levels of ${}_{\;\Lambda}^{10}\mathrm{Li}$ 
are very similar to those of ${}_{}^{9}\mathrm{Li}$ 
with \mbox{$T = \frac{3}{2}$}, 
\mbox{$J_{}^{\pi} = \frac{1}{2}_{}^{-}, \frac{3}{2}_{}^{-}$}, 
as seen in Fig.~\ref{f03}. 
We confirm that 
the ${}_{}^{9}\mathrm{Li}$ core state 
is slightly changed by the addition of the $\Lambda$ hyperon, 
and that the $\Lambda$ hyperon 
behaves as a single-particle motion in the nucleus~\cite{dov89} 
because the $\Lambda N$ interaction is rather weak. 
The results 
are supported by the calculated $\Lambda$-pickup spectroscopic factors 
of Eq.~(\ref{e10-form})\@. 
In Fig.~\ref{f03}, 
we also show the calculated spectroscopic factors $S_{\Lambda}^{}$ 
for the $\Lambda$ ground and two excited states; 
(a) \mbox{$(T, J_{}^{\pi}) = (\frac{3}{2}, 1_{}^{-})_{\mathrm{g.s.}}^{}$} 
at $0.0$ $\mathrm{MeV}$, 
(b) $(\frac{3}{2}, 1_{}^{-})_{5}^{}$ at $9.5$ $\mathrm{MeV}$, 
and 
(c) $(\frac{3}{2}, 1_{}^{-})_{10}^{}$ at $16.0$ $\mathrm{MeV}$. 
In the case of (a), 
we obtain 
\mbox{$S_{\Lambda}^{} \approx 1$} for the ${}_{}^{9}\mathrm{Li}$ ground state 
and \mbox{$S_{\Lambda}^{} \approx 0$} for other eigenstates. 
Similarly, 
in the case of (b), 
\mbox{$S_{\Lambda}^{} \approx 1$} for the ${}_{}^{9}\mathrm{Li}$ eigenstate. 
In the case of (c), 
\mbox{$S_{\Lambda}^{}$} has a large value for the two eigenstates, 
because these states are almost degenerate. 
\par
\begin{figure}[t]
  \includegraphics*[width=\linewidth]{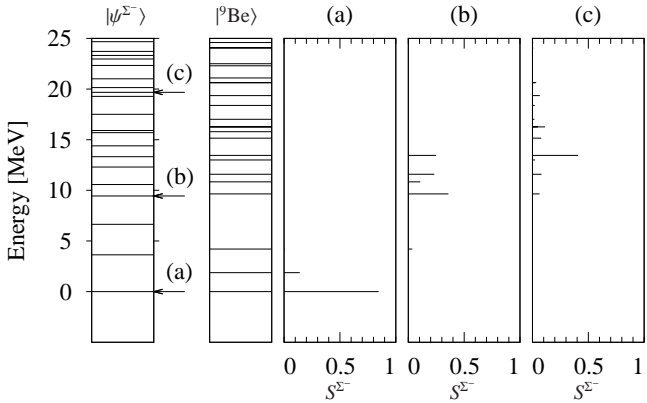}
  \caption%
  [Calculated energy spectra and spectroscopic factors]%
  {Calculated energy spectra 
   for $\Sigma$ eigenstates of ${}_{\;\Lambda}^{10}\mathrm{Li}$ 
   and eigenstates of ${}_{}^{9}\mathrm{Be}$. 
   $\Sigma_{}^{-}$-pickup spectroscopic factors for three eigenstates, 
   labeled by (a), (b), and (c), 
   are shown in each panel.} 
\label{f04}
\end{figure}
\begin{figure}[t]
  \includegraphics*[width=\linewidth]{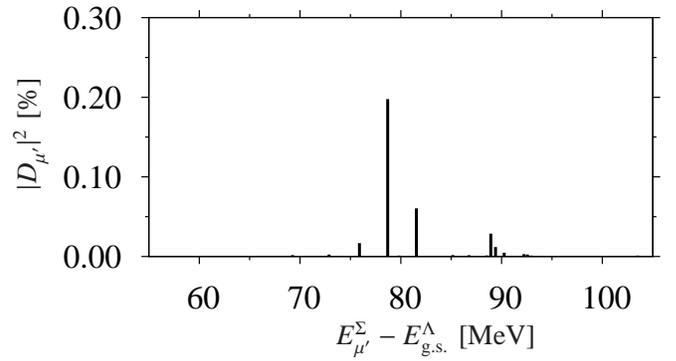}
  \caption%
  [$\Lambda$-$\Sigma$ coupling strengths]%
  {$\Lambda$-$\Sigma$ coupling strengths $|D_{\mu_{}^{\prime}}^{}|_{}^{2}$ 
   of the $\Sigma$ eigenstates 
   in the ground state of ${}_{\;\Lambda}^{10}\mathrm{Li}$.} 
\label{f05}
\end{figure}

In Fig.~\ref{f04}, 
we display the calculated energy spectra 
of $\Sigma$ eigenstates in ${}_{\;\Lambda}^{10}\mathrm{Li}$ 
and of ${}_{}^{9}\mathrm{Be}$ eigenstates 
with \mbox{$T_{}^{} = \frac{1}{2}, \frac{3}{2}, \frac{5}{2}$} 
and \mbox{$J_{}^{\pi} = \frac{1}{2}_{}^{-}, \frac{3}{2}_{}^{-}$}, 
together with the $\Sigma$-pickup spectroscopic factors $S_{\Sigma_{}^{-}}^{}$ 
for three states; 
(a) $(\frac{3}{2}, 1_{}^{-})_{\mathrm{g.s.}}^{}$ 
at $0.0$ $\mathrm{MeV}$, 
(b) $(\frac{3}{2}, 1_{}^{-})_{4}^{}$ at $9.4$ $\mathrm{MeV}$, 
and 
(c) $(\frac{3}{2}, 1_{}^{-})_{13}^{}$ at $19.7$ $\mathrm{MeV}$. 
The excited states of (b) and (c) 
are strongly coupled to the $\Lambda$ ground state 
in the energy regions of 
\mbox{$E_{\mathrm{g.s.}}^{\Sigma} - E_{\mathrm{g.s.}}^{\Lambda} \approx 80$} 
and $90$ $\mathrm{MeV}$, 
respectively, 
as we will mention in Fig.~\ref{f05}. 
The distributions of $S_{\Sigma_{}^{-}}^{}$ for (b) and (c) 
widely spread with the multi-configuration of ${}_{}^{9}\mathrm{Be}_{}^{\ast}$, 
as seen in Fig.~\ref{f04}. 
This implies that 
the $\Sigma$ hyperon 
has the ability of largely changing the core-nuclear configuration. 
\par
The calculated $\Lambda$-$\Sigma$ coupling strengths 
$|D_{\mu_{}^{\prime} }^{}|_{}^{2}$ 
between the $\Sigma$ eigenstates $| \psi_{\mu_{}^{\prime}}^{\Sigma} \rangle$ 
and the $\Lambda$ ground state $| \psi_{\mathrm{g.s.}}^{\Lambda} \rangle$ 
are shown in Fig.~\ref{f05}. 
It should be noticed that 
a contribution of the $\Sigma$ ground state 
$| \psi_{\mathrm{g.s.}}^{\Sigma} \rangle$ 
to the $\Sigma$-mixing of the ground state of ${}_{\;\Lambda}^{10}\mathrm{Li}$ 
is reduced to \mbox{$|D_{\mathrm{g.s.}}^{}|_{}^{2} = 0.002$} $\%$, 
whereas the several $\Sigma$ excited states in the 
\mbox{$E_{\mu_{}^{\prime}}^{\Sigma} - E_{\mathrm{g.s.}}^{\Lambda} \approx 80$} 
$\mathrm{MeV}$ region considerably contribute to the $\Sigma$-mixing. 
These contributions are coherently enhanced 
by the configuration mixing 
which is caused by the $\Sigma N$ interaction. 
It is shown that 
the nature of the $\Sigma$-nuclear states 
plays an important role in the $\Lambda$-$\Sigma$ coupling. 
\par
However, 
it should be noticed that 
the $\Sigma N$ interaction has still ambiguities. 
The values of radial integrals 
in the $\Sigma N$ \mbox{$T=\frac{3}{2}$} effective interaction, 
$\bar{V} = -1.107$ and $\Delta = 2.275$ $\mathrm{MeV}$ 
in NSC97e,f, 
lead to the attractive ${}_{}^{3}S_{1}^{}$ interaction of 
${}_{}^{33}V = \bar{V} + \frac{1}{4} \Delta = -0.538$ $\mathrm{MeV}$, 
as pointed out in Ref.~\cite{dab99}, 
whereas the \mbox{$T=\frac{3}{2}$}, ${}_{}^{3}S_{1}^{}$ interaction 
may be repulsive~\cite{fuj07y,rij08} 
because of recent suggestions of repulsive $\Sigma$-nuclear 
potentials~\cite{dab99,har91,nou02,fri07}. 
Further theoretical investigation are needed. 
%
%%%%%%%%%%%%%%%%%%%%%%%%%%%%%%%%%%%%%%%%%%%%%%%%%%%%%%%%%%%%%%%%%%%%%%%%%%%%%%%%
\section{Discussion}
\label{dis}
%%%%%%%%%%%%%%%%%%%%%%%%%%%%%%%%%%%%%%%%%%%%%%%%%%%%%%%%%%%%%%%%%%%%%%%%%%%%%%%%
%
\begin{figure}[t]
  \includegraphics*[width=\linewidth]{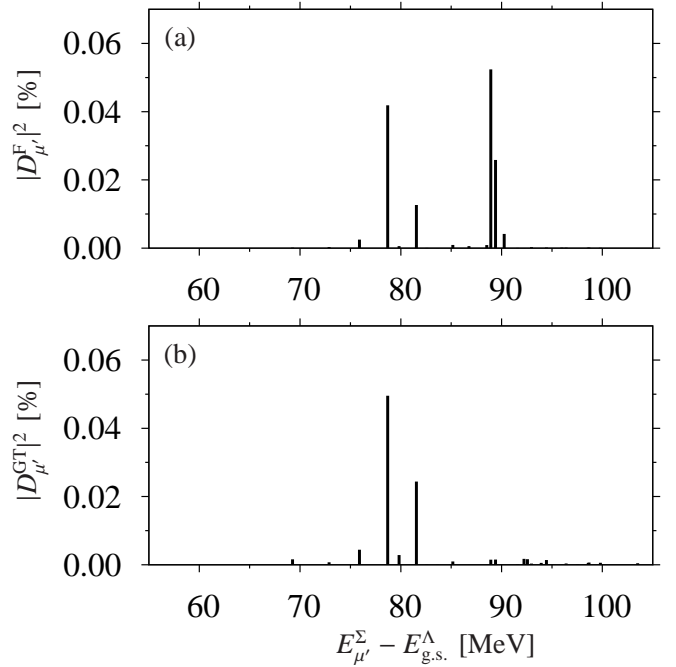}
  \caption%
  [Fermi and Gamow-Teller type coupling strengths]%
  {(a) 
   Fermi-type coupling strengths $|D_{\mu_{}^{\prime}}^{\mathrm{F}}|_{}^{2}$ 
   and 
   (b) 
   Gamow-Teller-type coupling strengths 
   $|D_{\mu_{}^{\prime}}^{\mathrm{GT}}|_{}^{2}$
   of the $\Sigma$ eigenstates 
   in the ground state of ${}_{\;\Lambda}^{10}\mathrm{Li}$.} 
\label{f06}
\end{figure}
\par
It is important to understand a mechanism of the $\Lambda$-$\Sigma$ coupling 
in neutron-rich nuclei microscopically. 
When a $\Lambda$ state in ${}_{\;\Lambda}^{10}\mathrm{Li}$ 
converts to a $\Sigma_{}^{-}$ state 
by the $\Lambda$-$\Sigma$ coupling interaction, 
the ${}_{}^{9}\mathrm{Li}$ core state 
changes into the ${}_{}^{9}\mathrm{Be}$ core state. 
In other words, 
the $\beta_{}^{-}$-transition, 
\mbox{${}_{}^{9}\mathrm{Li} \to {}_{}^{9}\mathrm{Be}$}, 
occurs in the core-nuclear state. 
Thus it is interesting 
to consider the $\beta$-transitions between the core-nuclear components 
of $\Lambda$ and $\Sigma$ eigenstates 
in order to investigate the strength distribution 
of the $\Lambda$-$\Sigma$ coupling. 
The two-body $\Lambda$-$\Sigma$ coupling interaction $V_{\Sigma\Lambda}^{}$ 
is approximately rewritten as 
\begin{gather}
\begin{aligned}[b]
  V_{\Sigma\Lambda}^{}
  &\simeq
  V_{\Sigma\Lambda}^{\mathrm{F}}
  +
  V_{\Sigma\Lambda}^{\mathrm{GT}}
\end{aligned}
\label{e01-dis}
\end{gather}
with 
\begin{gather}
\begin{aligned}[b]
  V_{\Sigma\Lambda}^{\mathrm{F}}
  &=
  V_{\mathrm{F}}^{}(r) 
  \bm{t}_{N}^{} \cdot \bm{\phi}_{\Sigma\Lambda}^{}
  ,
\end{aligned}
\label{e02-dis}
\\
\begin{aligned}[b]
  V_{\Sigma\Lambda}^{\mathrm{GT}}
  &=
  V_{\mathrm{GT}}^{}(r) 
  \mathinner{( \bm{\sigma}_{N}^{} \cdot \bm{\sigma}_{\Sigma\Lambda}^{} )}
  \bm{t}_{N}^{} \cdot \bm{\phi}_{\Sigma\Lambda}^{}
  ,
\end{aligned}
\label{e03-dis}
\end{gather}
where 
$\bm{t}_{N}^{}$ and $\bm{\sigma}_{N}^{} \bm{t}_{N}^{}$ 
denote the Fermi and Gamow-Teller $\beta$-transition operators for nucleons, 
respectively, 
and the operator 
$\bm{\phi}_{\Sigma\Lambda}^{}$ 
changes the $\Lambda$ hyperon into the $\Sigma$ hyperon, 
\begin{gather}
\begin{aligned}[b]
  | j_{\Sigma}^{} \rangle
  &=
  \bm{\phi}_{\Sigma\Lambda}^{}
  | j_{\Lambda}^{} \rangle
  ,
\end{aligned}
\label{e04-dis}
\end{gather}
and 
has tensorial rank-$0$ in the spin space and rank-$1$ in the isospin space; 
$\bm{\sigma}_{\Sigma\Lambda}^{}$ is the spin operator for a hyperon. 
Therefore, 
$V_{\Sigma\Lambda}^{\mathrm{F}}$ and $V_{\Sigma\Lambda}^{\mathrm{GT}}$ 
are regarded as the Fermi-type and Gamow-Teller-type coupling interactions, 
respectively. 
We stress that 
the $\Lambda$-$\Sigma$ coupling strengths $|D_{\mu_{}^{\prime}}^{}|_{}^{2}$ 
are extremely affected by strengths of these $\beta$-transitions 
between the core-nuclear states. 
We discuss 
which type interaction in $V_{\Sigma\Lambda}^{}$ 
contributes to the coupling strengths in the regions of 
\mbox{$E_{\mu_{}^{\prime}}^{\Sigma} - E_{\mathrm{g.s.}}^{\Lambda} \approx 80$} 
$\mathrm{MeV}$, 
as shown in Fig.~\ref{f05}. 
We can evaluate the strengths, 
\begin{gather}
\begin{aligned}[b]
  |D_{\mu_{}^{\prime}}^{\mathrm{F}}|_{}^{2}
  &=
  \left|
  \frac{ \langle \psi_{\mathrm{g.s.}}^{\Lambda}; T J |
         V_{\Sigma\Lambda}^{\mathrm{F}}
         | \psi_{\mu_{}^{\prime}}^{\Sigma}; T J \rangle }
       { E_{\mu_{}^{\prime}}^{\Sigma} - E_{\mathrm{g.s.}}^{\Lambda} }
  \right|_{}^{2}
  ,
\end{aligned}
\label{e05-dis}
\\
\begin{aligned}[b]
  |D_{\mu_{}^{\prime}}^{\mathrm{GT}}|_{}^{2}
  &=
  \left|
  \frac{ \langle \psi_{\mathrm{g.s.}}^{\Lambda}; T J |
         V_{\Sigma\Lambda}^{\mathrm{GT}}
         | \psi_{\mu_{}^{\prime}}^{\Sigma}; T J \rangle }
       { E_{\mu_{}^{\prime}}^{\Sigma} - E_{\mathrm{g.s.}}^{\Lambda} }
  \right|_{}^{2}
  ,
\end{aligned}
\label{e06-dis}
\end{gather}
using 
\mbox{$\bar{V}_{\Sigma\Lambda}^{} = 1.45$ $\mathrm{MeV}$} 
and 
\mbox{$\Delta_{\Sigma\Lambda}^{} = 3.04$ $\mathrm{MeV}$} 
in Table~\ref{t01}. 
The calculated strength distributions 
of the Fermi- and Gamow-Teller-type couplings 
are shown in Figs.~\ref{f06} (a) and (b), 
respectively. 
Comparing Fig.~\ref{f06} with Fig.~\ref{f05}, 
we recognize 
the Fermi and Gamow-Teller components 
coherently contribute to the $\Lambda$-$\Sigma$ coupling strengths 
in the energy region of 
\mbox{$E_{\mu_{}^{\prime}}^{\Sigma} - E_{\mathrm{g.s.}}^{\Lambda} \approx 80$} 
$\mathrm{MeV}$. 
The calculated $\Sigma$-mixing probability involving both couplings is 
\begin{gather}
\begin{aligned}[b]
  \sum_{\mu_{}^{\prime}}^{}
  \left|
  \frac{ \langle \psi_{\mathrm{g.s.}}^{\Lambda}; T J |
         \mathinner{(
           V_{\Sigma\Lambda}^{\mathrm{F}}
           +
           V_{\Sigma\Lambda}^{\mathrm{GT}}
         )}
         | \psi_{\mu_{}^{\prime}}^{\Sigma}; T J \rangle }
       { E_{\mu_{}^{\prime}}^{\Sigma} - E_{\mathrm{g.s.}}^{\Lambda} }
  \right|_{}^{2}
  =
  0.350 \; \%
  ,
\end{aligned}
\label{e07-dis}
\end{gather}
which is close to the full calculated probability 
\mbox{$P_{\Sigma}^{} = 0.345$} $\%$, 
while 
\mbox{$\sum_{\mu_{}^{\prime}}^{} |D_{\mu_{}^{\prime}}^{\mathrm{F}}|_{}^{2}
      = 0.144$} $\%$ 
is obtained for the Fermi type 
and 
\mbox{$\sum_{\mu_{}^{\prime}}^{} |D_{\mu_{}^{\prime}}^{\mathrm{GT}}|_{}^{2} 
      = 0.098$} $\%$
for the Gamow-Teller type. 
\par
The Fermi-type operator 
\mbox{$\bm{t}_{N}^{} \cdot \bm{\phi}_{\Sigma\Lambda}^{}$} 
does not change the core-nuclear states in 
$| \psi_{\mathrm{g.s.}}^{\Lambda} \rangle$, 
which is equivalent to the Fermi $\beta$-transition, 
$| ({}_{}^{A}Z) \nu T T_{z}^{} J M \rangle$ 
$\to$ 
$| ({}_{}^{A}Z{\pm}1) \nu T \, T_{z}^{}{\pm}1 \, J M \rangle$. 
Note that this transition can change only $T_{z}^{}$. 
By the Fermi $\beta$-transition 
from the ${}_{}^{9}\mathrm{Li}$ \mbox{$T = \frac{3}{2}$} ground state, 
the ${}_{}^{9}\mathrm{Be}_{}^{\ast}$ \mbox{$T = \frac{3}{2}$} excited state 
is populated around 
\mbox{$E_{x}^{}({}_{}^{9}\mathrm{Be}_{}^{*}) \simeq 14$} $\mathrm{MeV}$, 
measured from 
the ${}_{}^{9}\mathrm{Be}$ \mbox{$T = \frac{1}{2}$} ground state. 
If the weak-coupling limit in the $\Sigma N$ interaction 
works well like the $\Lambda N$ interaction, 
the coupling is concentrated on the only one state. 
As a result, 
one peak arises at 
\mbox{$E_{\mu_{}^{\prime}}^{\Sigma} - E_{\mathrm{g.s.}}^{\Lambda} \approx 80$} 
$\mathrm{MeV}$ 
in the coupling strength. 
However, 
the calculated strength distribution of 
$|D_{\mu_{}^{\prime}}^{\mathrm{F}}|_{}^{2}$ 
widely spreads to the energy regions of 
\mbox{$E_{\mu_{}^{\prime}}^{\Sigma} - E_{\mathrm{g.s.}}^{\Lambda} = 75$}--$80$ 
and about $90$ $\mathrm{MeV}$, 
as shown in Fig.~\ref{f06} (a). 
This means that 
the $\Sigma$ hyperon changes a nuclear configuration mixing 
due to the strong spin-isospin dependence in the $\Sigma N$ interaction, 
leading to the coherence of the $\Lambda$-$\Sigma$ coupling. 
\par
In order to see the potentiality of the neutron-rich nuclei clearly, 
we discuss 
why the energy shift for ${}_{\;\Lambda}^{10}\mathrm{Li}$ 
is about 3 times larger than that for ${}_{\Lambda}^{7}\mathrm{Li}$. 
The enhancement of the $\Sigma$-mixing probabilities 
in neutron-rich $\Lambda$-hypernuclei 
is mainly due to the Fermi-type coupling interaction 
$V_{\Sigma\Lambda}^{\mathrm{F}}$, 
which might correspond to the coherent $\Lambda$-$\Sigma$ coupling 
suggested in Refs.~\cite{swe00,aka00}. 
When we use the weak-coupling limit for simplicity, 
we consider a matrix element of $V_{\Sigma\Lambda}^{\mathrm{F}}$ 
between $\Lambda$ and $\Sigma$ states, 
\begin{gather}
\begin{aligned}[b]
  \langle V_{\Sigma\Lambda}^{\mathrm{\mathrm{F}}} \rangle
  &=
  \langle T_{N}^{}{=}T J_{N}^{}, j_{\Sigma}^{}; T J |
  V_{\Sigma\Lambda}^{\mathrm{\mathrm{F}}}
  | T_{N}^{}{=}T J_{N}^{}, j_{\Lambda}^{}; T J \rangle
  ,
\end{aligned}
\label{e08-dis}
\end{gather}
where 
the $\Lambda$ and $\Sigma$ states 
must have the same core-nuclear state 
$| T_{N}^{} J_{N}^{} \rangle$. 
If we assume that 
\mbox{$V_{\mathrm{F}}^{}(r) = \bar{V}_{\mathrm{F}}^{}$}, 
we obtain 
\begin{gather}
\begin{aligned}[b]
  \langle V_{\Sigma\Lambda}^{\mathrm{\mathrm{F}}} \rangle
  &=
  \bar{V}
  \sqrt{ \frac{ 4 T \mathinner{(T+1)} }{ 3 } }
\end{aligned}
\label{e09-dis}
\end{gather}
and find that 
the Fermi-type coupling strength $|D_{\mu_{}^{\prime}}^{\mathrm{F}}|_{}^{2}$ 
is proportional to $T \mathinner{(T+1)}$. 
Therefore, 
the Fermi-type coupling strengths 
play the important role in the $\Sigma$-mixing probability 
of neutron-rich ${}_{\;\Lambda}^{10}\mathrm{Li}$ eigenstates 
because of \mbox{$T = \frac{3}{2}$}. 
On the other hand, 
in ${}_{\Lambda}^{7}\mathrm{Li}$ \mbox{$T = 0$} states, 
the Fermi-type coupling strengths vanish 
whereas 
the $\Sigma$-mixing probability 
is mainly caused by the Gamow-Teller-type coupling strengths. 
\par
In the Gamow-Teller transitions for ordinary nuclei, 
the sum rule~\cite{ike63} 
\begin{gather}
\begin{aligned}[b]
  \sum B(\mathrm{GT-}) - \sum B(\mathrm{GT+})
  &=
  3 \mathinner{( N - Z )}
\end{aligned}
\label{}
\end{gather}
is well known as a model independent formula, 
where 
$B(\mathrm{GT\mp})$ 
is a strength of the Gamow-Teller $\beta_{}^{\mp}$-transition, 
$| {}_{}^{A}Z \rangle \to | {}_{}^{A}Z{\pm}1 \rangle$. 
In general, 
$\sum B(\mathrm{GT+})$ becomes smaller as neutron-excess grows larger, 
leading to 
$\sum B(\mathrm{GT-}) \approx 3 \mathinner{( N - Z )}$. 
Therefore, 
the Gamow-Teller-type coupling 
is very important 
in $\Lambda$-hypernuclei with large neutron excess. 
%
%%%%%%%%%%%%%%%%%%%%%%%%%%%%%%%%%%%%%%%%%%%%%%%%%%%%%%%%%%%%%%%%%%%%%%%%%%%%%%%%
\section{Summary and conclusion}
\label{conc}
%%%%%%%%%%%%%%%%%%%%%%%%%%%%%%%%%%%%%%%%%%%%%%%%%%%%%%%%%%%%%%%%%%%%%%%%%%%%%%%%
%
We have investigated the structure 
of the neutron-rich ${}_{\;\Lambda}^{10}\mathrm{Li}$ hypernucleus, 
in shell-model calculations considering the $\Lambda$-$\Sigma$ coupling 
in the perturbation theory. 
We have found that 
the $\Sigma$-mixing probabilities and the energy shifts 
of ${}_{\;\Lambda}^{10}\mathrm{Li}$ eigenstates 
are coherently enhanced 
by the $\Lambda$-$\Sigma$ coupling configurations 
in the neutron-rich nucleus. 
We have argued the effects of the $\Lambda$-$\Sigma$ coupling interaction 
in terms of the $\beta$-transitions for the core-nuclear states. 
The reasons why the $\Sigma$-mixing probabilities are enhanced 
are summarized as follows: 
(i) 
The multi-configuration $\Sigma$ excited states 
can be strongly coupled with the $\Lambda$ ground state 
with the help of the $\Sigma N$ interaction. 
(ii) 
These strong $\Lambda$-$\Sigma$ couplings 
are coherently enhanced 
by the Fermi- and Gamow-Teller-type coupling components. 
(iii) 
The Fermi-type coupling 
becomes more effective in the neutron-rich environment 
increasing as $T \mathinner{(T+1)}$. 
\par
In conclusion, 
we have found that 
the $\Sigma$-mixing probability is about $0.34$ $\%$ 
and the energy shift is about $0.28$ $\mathrm{MeV}$ 
for the neutron-rich ${}_{\;\Lambda}^{10}\mathrm{Li}$ 
$1_{\mathrm{g.s.}}^{-}$ ground state, 
which is about 3 times larger than 
that for ${}_{\Lambda}^{7}\mathrm{Li}$. 
This is the first estimation of the $\Sigma$-mixing 
in the neutron-rich hypernuclei 
by microscopic shell-model calculations. 
%
%%%%%%%%%%%%%%%%%%%%%%%%%%%%%%%%%%%%%%%%%%%%%%%%%%%%%%%%%%%%%%%%%%%%%%%%%%%%%%%%
%
\begin{acknowledgments}
The authors are obliged to K. Muto for valuable discussion and useful comments. 
They are pleased to acknowledge D.~J. Millener 
for providing the $\Sigma N$ effective interaction in $p$ shell 
based on NSC97e,f 
and 
for valuable discussion and comments. 
This work is supported by Grant-in-Aid for Scientific Research 
on Priority Areas (Nos.~17070002 and 20028010). 
\end{acknowledgments}
%
%%%%%%%%%%%%%%%%%%%%%%%%%%%%%%%%%%%%%%%%%%%%%%%%%%%%%%%%%%%%%%%%%%%%%%%%%%%%%%%%
%
%%% REFERENCES %%%%%%%%%%%%%%%%%%%%%%%%%%%%%%%%%%%%%%%%%%%%%%%%%%%%%%%%%%%%%%%%%
%

%
%%%%%%%%%%%%%%%%%%%%%%%%%%%%%%%%%%%%%%%%%%%%%%%%%%%%%%%%%%%%%%%%%%%%%%%%%%%%%%%%
%
\end{document}